\documentclass[aps,pra,twocolumn,showpacs,groupedaddress]{revtex4}
\usepackage{color}
\usepackage{graphicx}
\usepackage{dcolumn}
\usepackage{bm}
\usepackage{amssymb}
\usepackage{hyperref}

\def\etl{$et~al.~$}

\begin{document}

\title{Saddle-Node Bifurcation of Periodic Orbit Route to Hidden Attractors in Nonlinear Dynamical Systems}
\author{Suresh Kumarasamy$^1$, Malay Banerjee$^2$, Vaibhav Varshney$^3$, Manish Dev Shrimali$^{4}$, Nikolay V. Kuznetsov$^{5}$, Awadhesh Prasad$^{6}$}
\affiliation{$^{1}$Centre for Nonlinear Systems, Chennai Institute of Technology, Chennai-600 069, Tamilnadu, India.}
\affiliation{$^{2}$Department of Mathematics and Statistics, Indian Institute of Technology Kanpur, Kanpur, 208016, Uttar Pradesh, India.}
\affiliation{$^{3}$Department of Physics, University of Engineering and Management, Jaipur, Rajasthan 303807, India.}
\affiliation{$^{4}$ Department of Physics, Central University of Rajasthan, Ajmer 305 817, Rajasthan, India.}
\affiliation{$^{5}$Department of Applied Cybernetics, Saint-Petersburg State University, Saint-Petersburg, Russia.}
\affiliation{$^{6}$Department of Physics and Astrophysics, University of Delhi, Delhi 110007, India.}

\begin{abstract}
Hidden attractors are present in many nonlinear dynamical systems and are not associated with equilibria, making them difficult to locate. Recent studies have demonstrated methods of locating hidden attractors, but the route to these attractors is still not fully understood. In this letter, we present the route to hidden attractors in systems with stable equilibrium points and in systems without any equilibrium points. We show that hidden attractors emerge as a result of the saddle-node bifurcation of stable and unstable periodic orbits. Real-time hardware experiments were performed to demonstrate the existence of hidden attractors in these systems. Despite the difficulties in identifying the suitable initial conditions from the appropriate basin of attraction, we performed experiments to detect hidden attractors in nonlinear electronic circuits. Our results provide new insights into the generation of hidden attractors in nonlinear dynamical systems.
\end{abstract}
\maketitle

The study of nonlinear and complex systems often involves the emergence of oscillations and even complex oscillations. Understanding the emergence of these nonlinear oscillations has been of utmost importance since the early 20th century. In nonlinear systems, the self-excited attractors can be easily identified by locating the trajectory emerges from the neighborhood of an unstable equilibrium. When the initial condition is selected near the equilibrium point, the trajectory will eventually settle in the self-excited attractor after a period of transient dynamics.

However, this approach is not directly applicable to multi-stable and mega-stable dynamical systems, which contain multiple coexisting attractors \cite{dawid1,ANP}. These types of systems can be found in climate models, the human brain, slowly rotating pendula \cite{tkpr}, and ecological systems \cite{Scheffer}. Recent studies have revealed a class of multi-stable attractors called hidden attractors, which are difficult to identify and locate \cite{Leonov,dawid,jafaripp}. The concept of hidden attractors was first introduced in connection with Hilbert’s 16th problem formulated in 1900 \cite{Hilbert, Kuznetsov2020}. Kuznetsov et al. reported the chaotic hidden attractor in Chua's circuit \cite{Kuznetsov2010,Kuznetsov2023}. This research has since led to further efforts to locate hidden attractors \cite{Leonov2013,Leonov2011,jafri}.

The basins of attraction of hidden attractors do not touch the unstable manifold of saddle fixed points and are located away from these points, making them difficult to detect using standard computational methods. However, there have been some successful efforts to locate hidden attractors using perpetual points \cite{Leonov, appp, jafaripp}. Hidden attractors can be generated through boundary crisis of an existing chaotic saddle \cite{Yan} or through flattening of trajectories coming from infinity that are unrelated to any equilibrium states \cite{Leonov}. Hidden global attractors cannot  exist in the Euclidean phase space, where the global attractor always contains at least one equilibrium state, but can exist in a cylindrical phase space \cite{Arnold1,nkv}. However, the exact methodology to determine the route to a hidden attractor in dynamical systems remains unknown.

\begin{figure}[h]  
 \centering
 {\scalebox{0.35}{\includegraphics{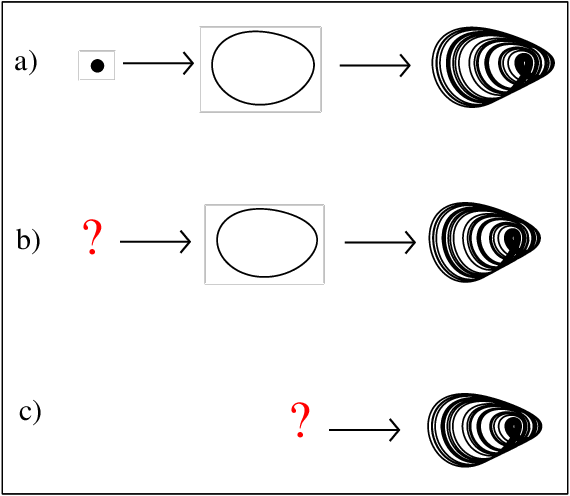}}}
 \caption{Schematic diagram for the creation of an attractor.}
 \label{fig:sch} 
 \end{figure}

\par Let us now consider case (a) in the schematic diagram in Figure \ref{fig:sch}. When a system is unable to reach an equilibrium point due to a change in stability, it can result in periodic oscillations through bifurcations, such as Hopf bifurcation. As the parameter changes, successive bifurcations may lead to chaotic dynamics in the system. The question of importance is, if the only equilibrium point is remains stable under the bifurcation, what are the successive bifurcations or routes to the creation of periodic attractors, also known as hidden attractors? Another class of dynamic systems is where there are no equilibrium points (case (c)), and in such cases, hidden attractors can seemingly appear from nowhere. The intriguing question is, how do hidden attractors arise from nothing, and what are the routes for the creation of such hidden attractors in systems with no equilibrium points (as shown in case (b) in Figure \ref{fig:sch})? Currently, the routes to the hidden attractors are not well understood.

To answer this intriguing question, we have found that hidden attractors appear as a result of saddle-node bifurcation of periodic and unstable periodic orbits in certain dynamic systems. We will present two cases of systems where hidden attractors appear through saddle-node bifurcation of periodic orbits. Case (i) involves systems with one stable equilibrium point, while case (ii) involves a system with no equilibrium points.

\begin{figure}[h]
  \centering
   {\scalebox{0.15}{\includegraphics{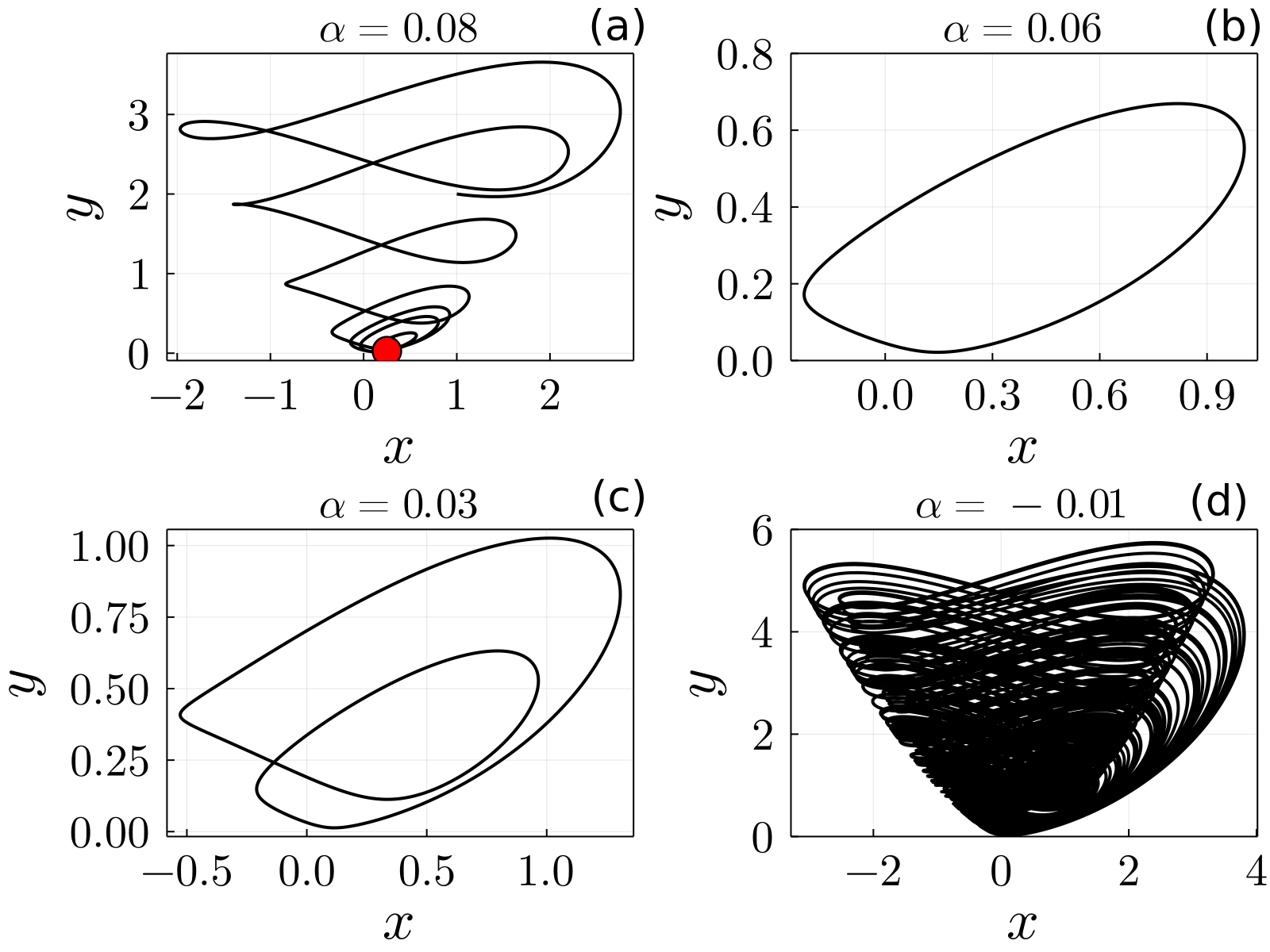}}}
 \caption{(a) Fixed point attractor along with transient trajectory ($\alpha=0.08$), (b) period-1 ($\alpha=0.06$) (c) period-2 ($\alpha=0.03$) and chaotic attractor ($\alpha=-0.01$) for system, Eq. (\ref{wang1}). }
 \label{fig:wang2}
 \end{figure}

\noindent
The letter is organized as follows: First, we present the emergence of hidden attractors in a nonlinear system with one stable fixed point, as introduced by Wang \etl \cite{wang}. We illustrate the process of generating hidden attractors and the basin of attraction for these attractors. Next, we demonstrate the generation of hidden attractors in a nonlinear system without fixed points. Finally, we provide a summary and conclusion of our work.\\
To begin, let us examine an instance of a nonlinear system that features a stable equilibrium point. This particular system was originally introduced by Wang et al. \cite{wang,wang1}.
\begin{eqnarray}
	\nonumber
        \dot{x}&=&yz+\alpha\\
	\nonumber
        \dot{y}&=& x^2-y\\
        \dot{z}&=& 1-4x,
\label{wang1}     
\end{eqnarray}
where $\alpha$ is a parameter. 
The system shows the period doubling route to chaos as we decrease the parameter $\alpha$. Stability analysis shows that the system has one stable fixed point at $(1/4,1/16,-16\alpha)$, when the parameter $\alpha>0$. For the negative $\alpha$ the fixed point is unstable. At a higher value of $\alpha \gtrsim 0.065$ the system has only a stable fixed point attractor and no other existing attractor  (as shown in Fig. \ref{fig:wang2}). When the parameter $\alpha<0$ the system has no attractor. The system shows period doubling route to chaos in the parameter interval $\alpha \in (0,0.065)$. 
 
As the parameter $\alpha$  decreased from $0.065$ a period-one limit cycle is created out of the blue sky. The period-one limit cycle is a hidden attractor because the basin contain no fixed point. The basin boundary does not intersect with any fixed point of the system.  On the further decrease of $\alpha$ the period-doubling bifurcation leads to chaotic hidden oscillations. Fig. \ref{fig:wang2} shows various dynamics observed in the system for different values of the parameter $\alpha$.\\

\begin{figure}[ht]
\centering
\vskip.5cm
 {\scalebox{0.035}{\includegraphics{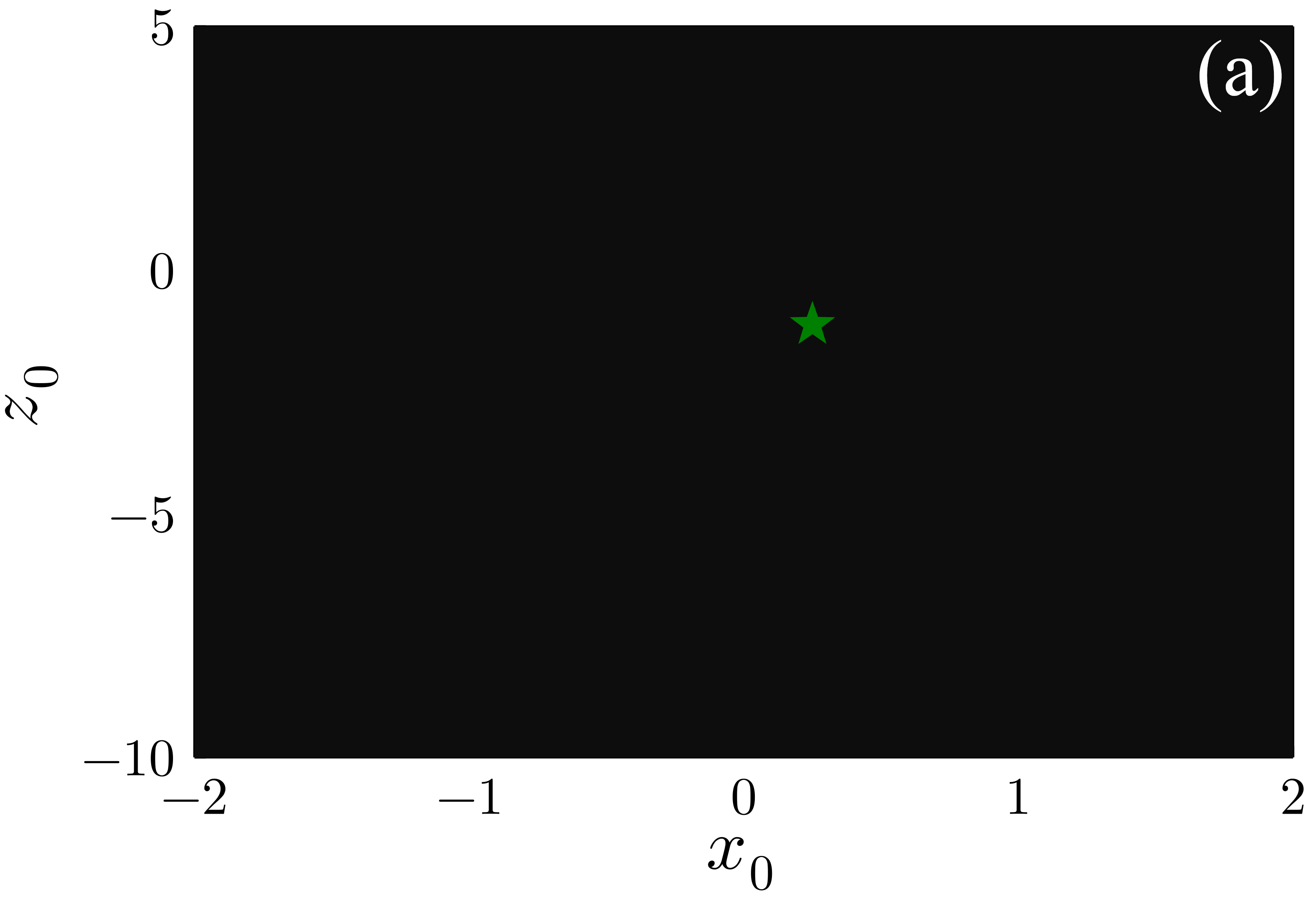}}}
 {\scalebox{0.035}{\includegraphics{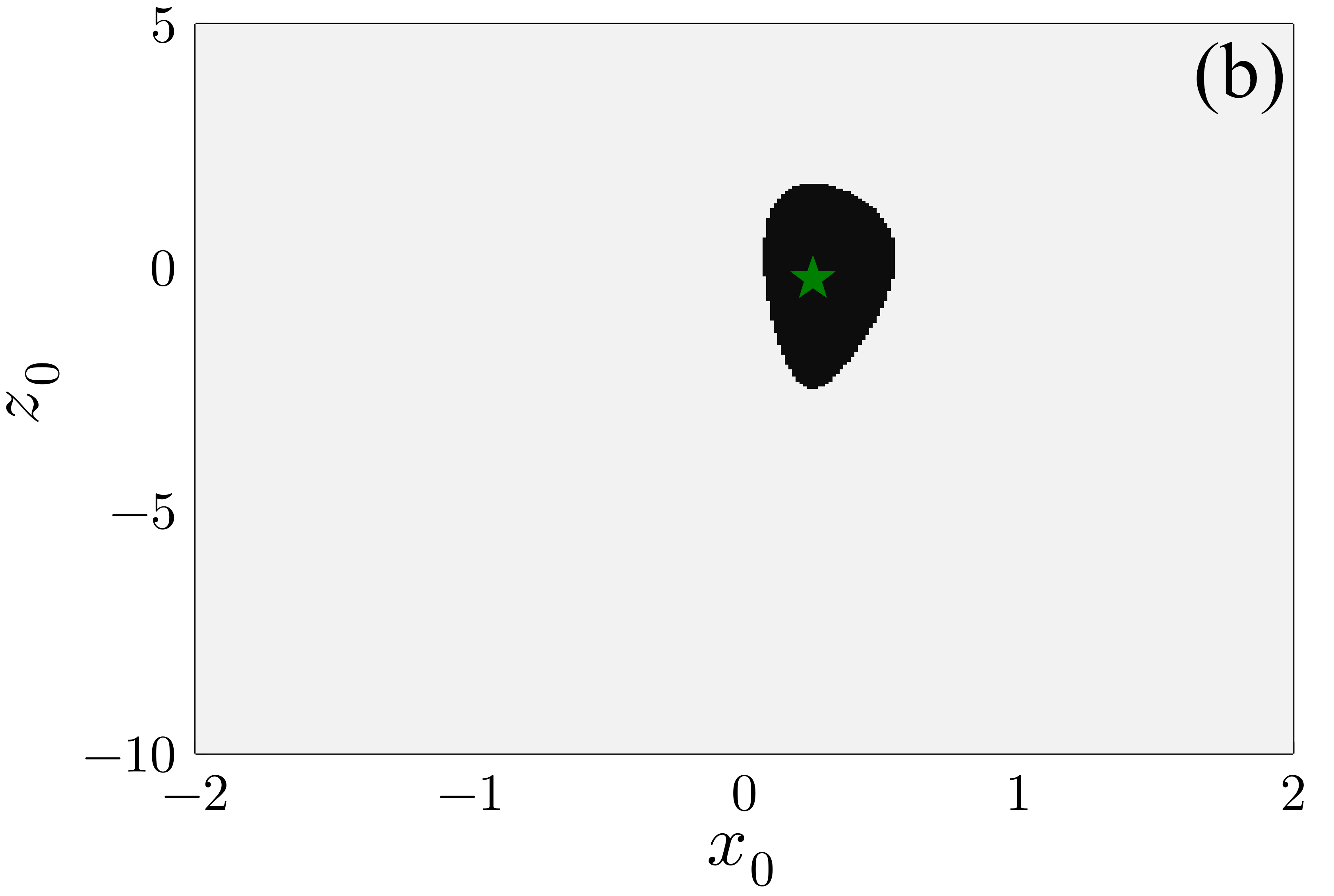}}}
 {\scalebox{0.035}{\includegraphics{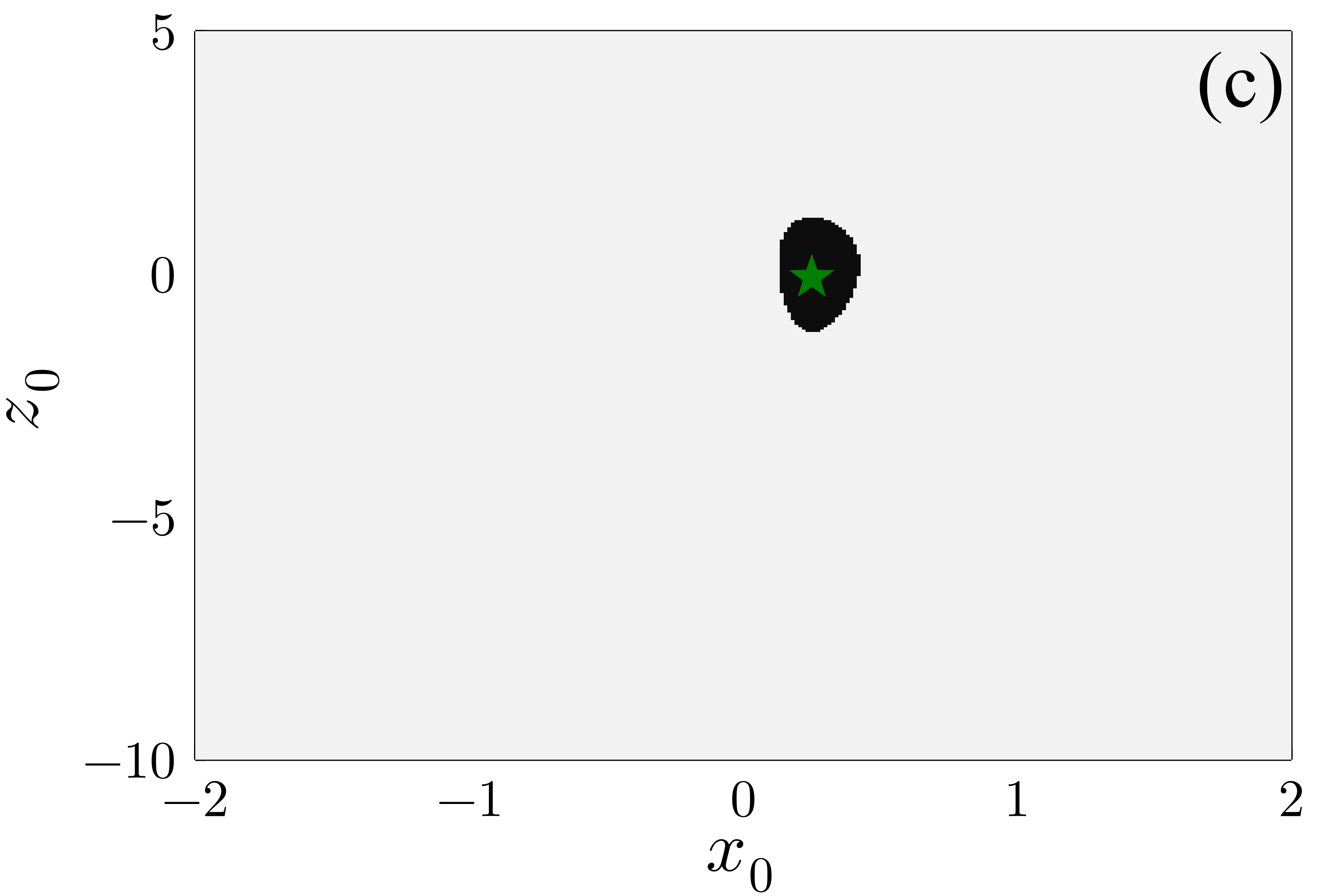}}}
 {\scalebox{0.18}{\includegraphics{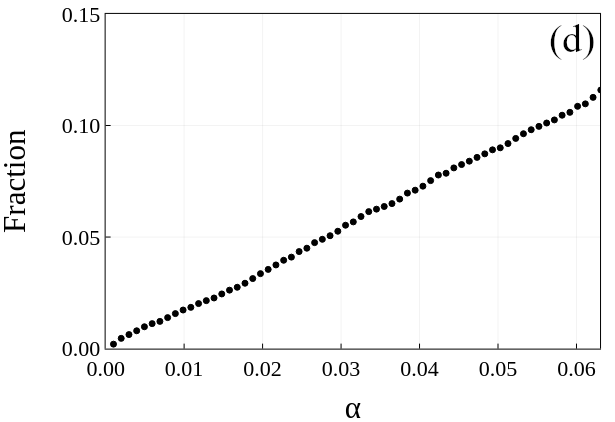}}}
 
 \caption{ Basin of attraction of initial conditions for (a) $\alpha=0.8$, (b) $\alpha=0.0148$ and (c) $\alpha=0.0045$ and (d) fraction of  locally attractive stationary set for system, Eq. (\ref{wang1}). }
 \label{basin}
 \end{figure}

\emph{Role of unstable periodic orbits in the creation of hidden attractor.} Now, let us explore how hidden attractors are created in systems with only one fixed point. For the parameter $\alpha \gtrsim 0.065$, the system has a globally stable (globally attracting) stationary set, as shown in Fig. \ref{basin}(a). The green star in the figure represents the stable fixed point attractor. In this parameter range, initial conditions starting from anywhere in the state space converge to this globally attracting fixed point.

As we decrease the parameter to $\alpha<0.065$, the global stability of the stationary set (fixed point) is violated by the appearance of boundaries due to global bifurcations away from the vicinity of the stationary set. This global bifurcation creates hidden boundaries for the global stability of the system. If an attractor is born via such a non-local bifurcation that causes the loss of global stability, then the attractor is hidden because the basin of its attraction is separated from the locally attractive stationary set.

\begin{figure}[ht]
\centering
\vskip.5cm
 {\scalebox{0.035}{\includegraphics{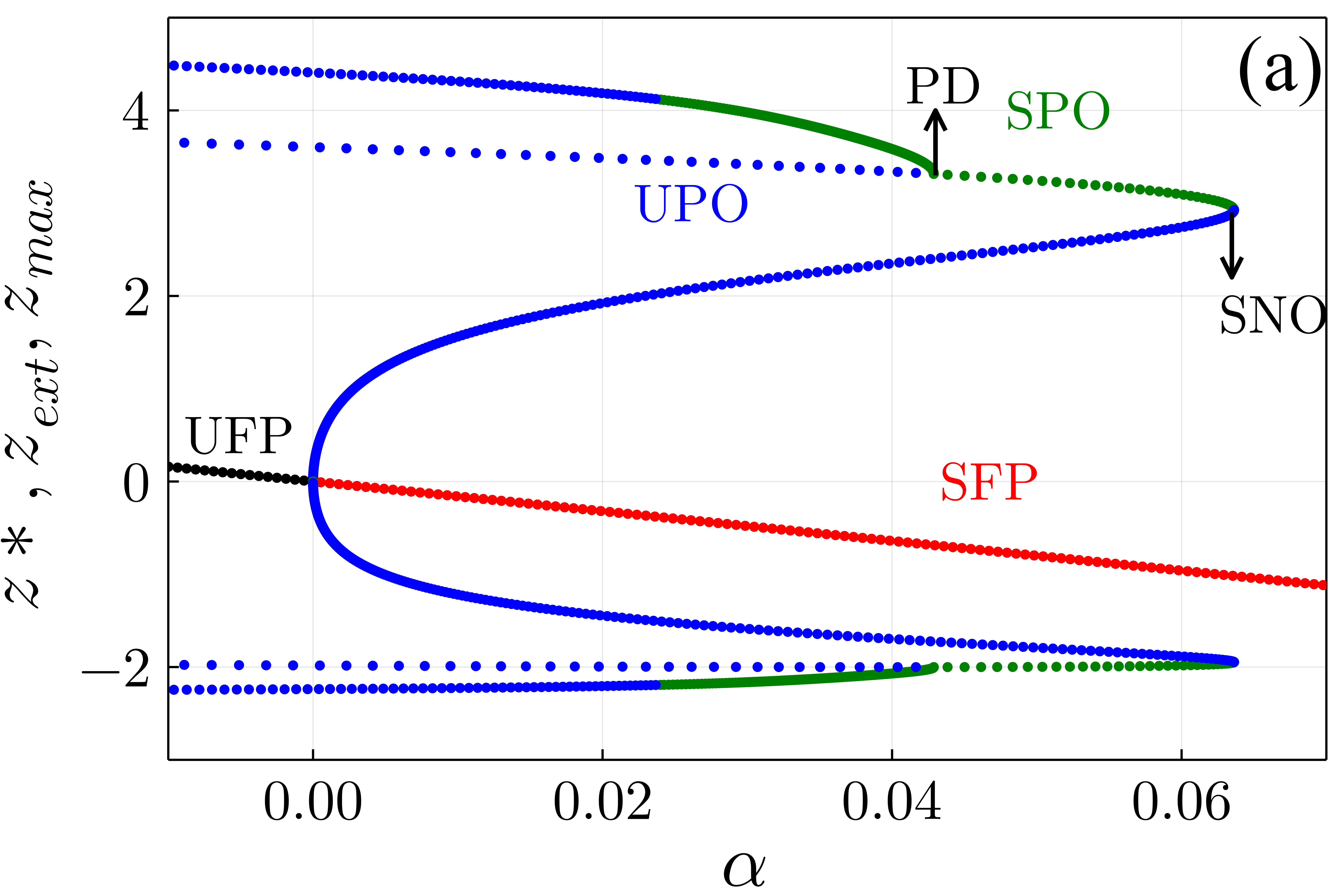}}}\\
  {\scalebox{0.035}{\includegraphics{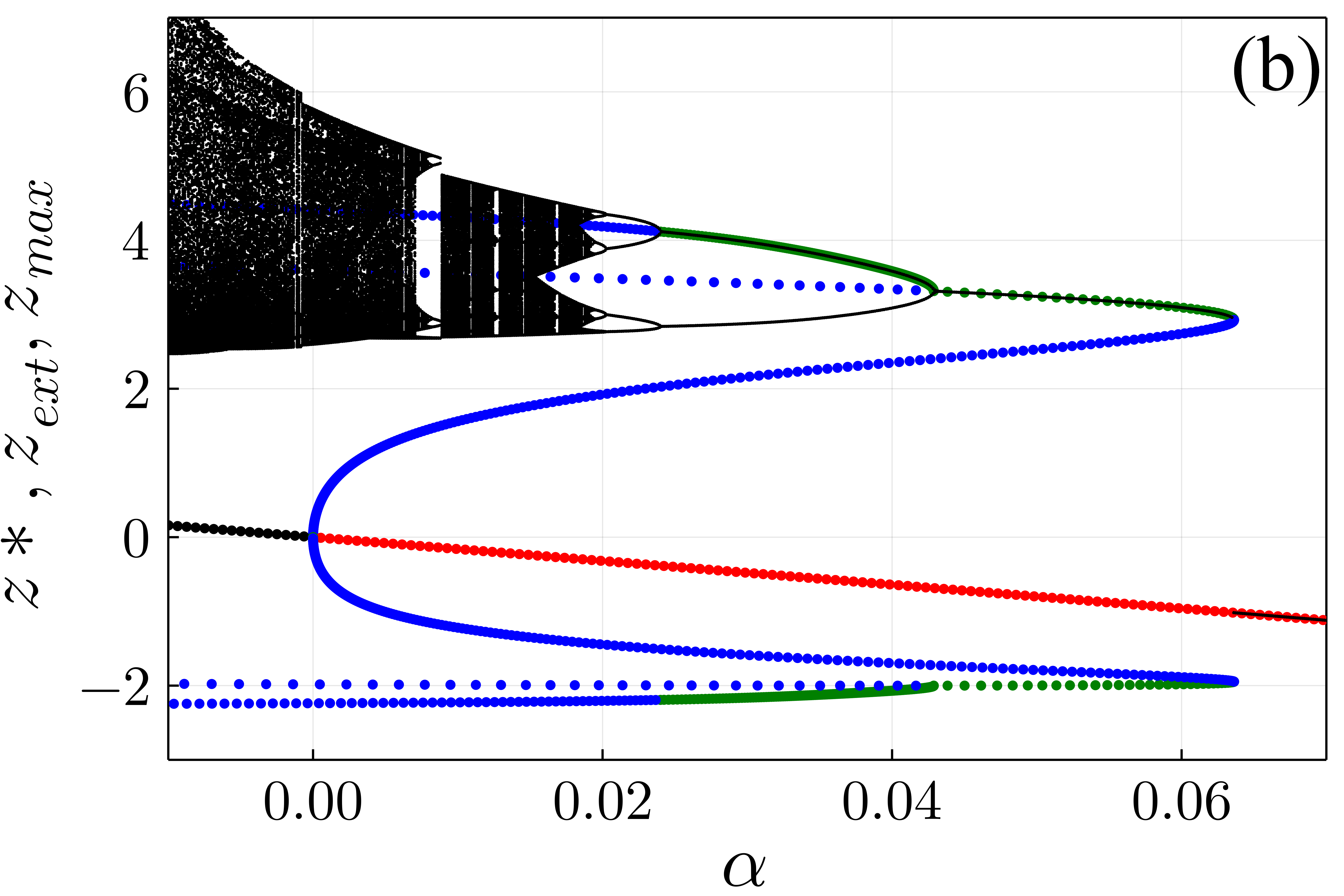}}}
 \caption{ (a) Fixed points and continuation diagram for the system Eq. (\ref{wang1}) as a function of parameter $\alpha$. SNO denotes the point of saddle node bifurcation on orbits. SPO and UPO are the stable and unstable periodic orbits. PD is the periodic doubling bifurcation points. (b) The bifurcation diagram overlapped with the continuation diagram. }
 \label{s1cont}
 \end{figure}

Fig. \ref{basin}(b) shows the basins of attraction of initial conditions $x_0\in(-2,2)$ and $z_0\in(-10,5)$ of Eq. (\ref{wang1}) for $\alpha=0.0148$ when projected on $xz$-plane. The original fixed point of the system remains attractive and has a basin of attraction around it, represented by the black-shaded region in the plot. The light gray region outside represents the basin of attraction corresponding to the hidden attractor.
To understand the global bifurcation that leads to the birth of hidden attractors, we plotted the fixed points and their stability, also known as the continuation diagram (Fig. \ref{s1cont}). As the parameter $\alpha$ decreases to $0.065$, the system exhibits a saddle-node bifurcation of periodic orbits (marked as SNO). The green dotted line in the plot represents the extrema of the stable periodic orbit, and the blue dotted line represents the extrema of unstable periodic orbit. The red dotted line in Fig. \ref{s1cont}(a) shows the stable fixed points, and the black dotted lines show the unstable fixed points. The stable unstable periodic orbits are created with the help of $XPPAUT$ \cite{xppaut}.  This global saddle-node bifurcation on the orbit annihilates the global stability of the attracting set and creates the basin boundary. The basin boundary is separated by the unstable periodic orbits, which act as a separatrix. As we decrease the parameter $\alpha$ towards zero, the area of the basin containing the attracting fixed point shrinks (Fig. \ref{basin}(c) for $\alpha=0.0045$), which can be understood from the maxima of the unstable periodic orbits. The plot in Fig. \ref{basin}(d) shows the fraction of the set of initial conditions that lead to the locally stable attraction. The fraction decreases as we approach the parameter $\alpha \to 0$. We can see from Fig. \ref{s1cont}(a) that the maxima and minima of the branch of the period one unstable limit cycle approach each other. When $\alpha=0$, there is no width between these two branches. Essentially, these two branches collide with each other, and the stability of the fixed point changes from an attracting fixed point to a repelling (unstable) fixed point. In other words, we have a subcritical Hopf bifurcation at $\alpha=0$. While decreasing the parameter to $\alpha<0$, the stability of the equilibrium point of the system becomes unstable.
\par The above results clearly show that the system has local bifurcation (i) At $\alpha=0$, the system exhibits subcritical Hopf bifurcation, and global bifurcation (ii) at $\alpha=0.065$, the system shows saddle-node bifurcation of periodic orbit. After the saddle-node bifurcation, further decreasing the parameter leads to successive periodic doubling bifurcation of hidden attractors, ultimately leading to hidden chaotic attractors in the system. Note that the hidden attractors (basin of hidden attractors) created through the saddle-node bifurcation of orbits continue to exist in the system irrespective of other local bifurcations in the system. For example, the system has a subcritical Hopf bifurcation at $\alpha=0$, and this local bifurcation does not affect the existence of hidden attractors created through saddle-node bifurcation. The birth of hidden attractors through this saddle-node bifurcation of orbits is found in other systems with one stable fixed point.

\begin{figure}[h!]
 {\scalebox{0.035}{\includegraphics[angle=0]{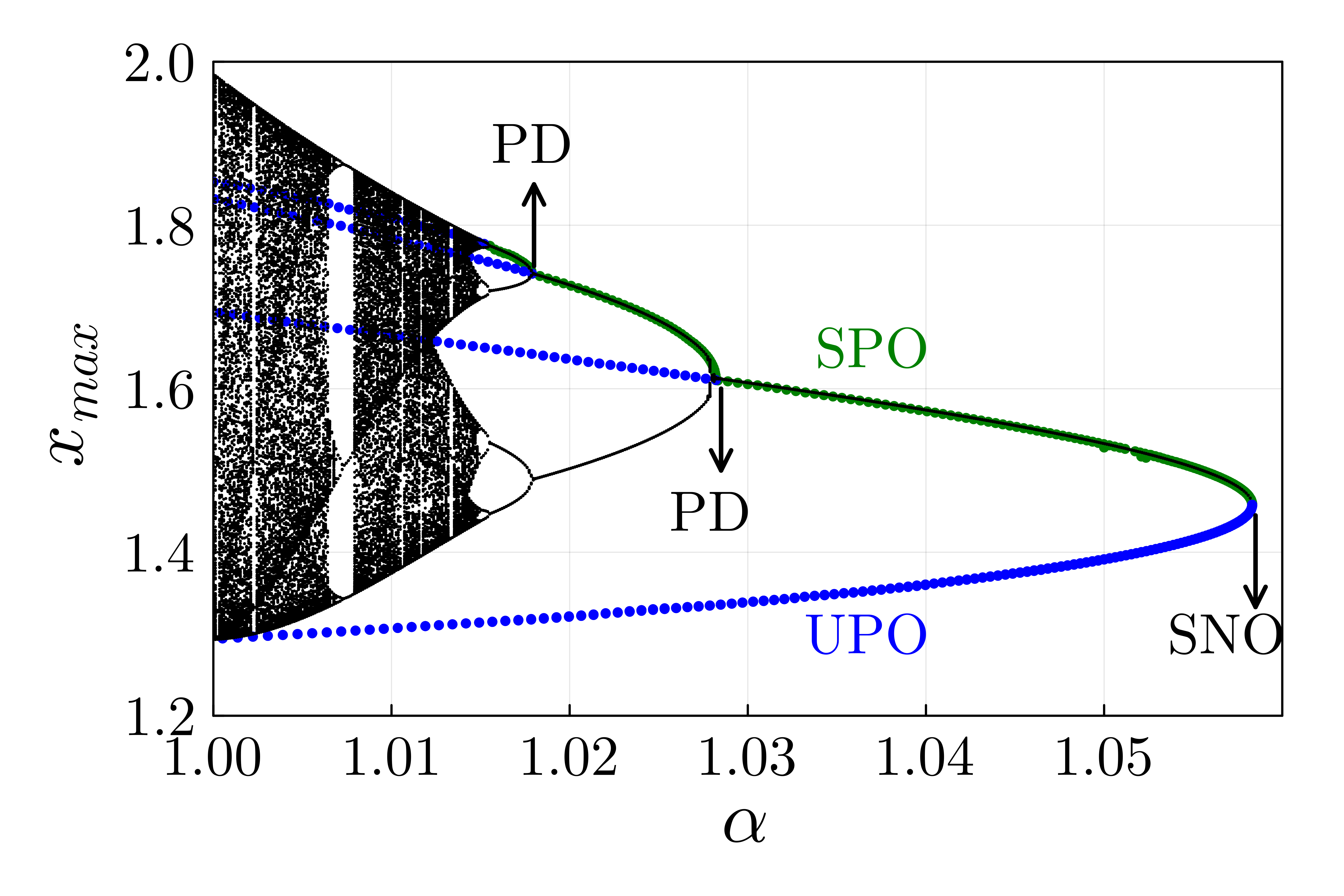}}} 
	\caption{ (a) Bifurcation diagram as a function of parameter $\alpha$ for system, Eq. (\ref{wang}). SNO denotes the point of saddle node bifurcation on orbits. SPO and UPO are the stable and unstable periodic orbits. PD is the periodic doubling bifurcation points.}
 \label{fig:wang}
 \end{figure}

\begin{figure}
 {\scalebox{0.13}{\includegraphics[angle=0]{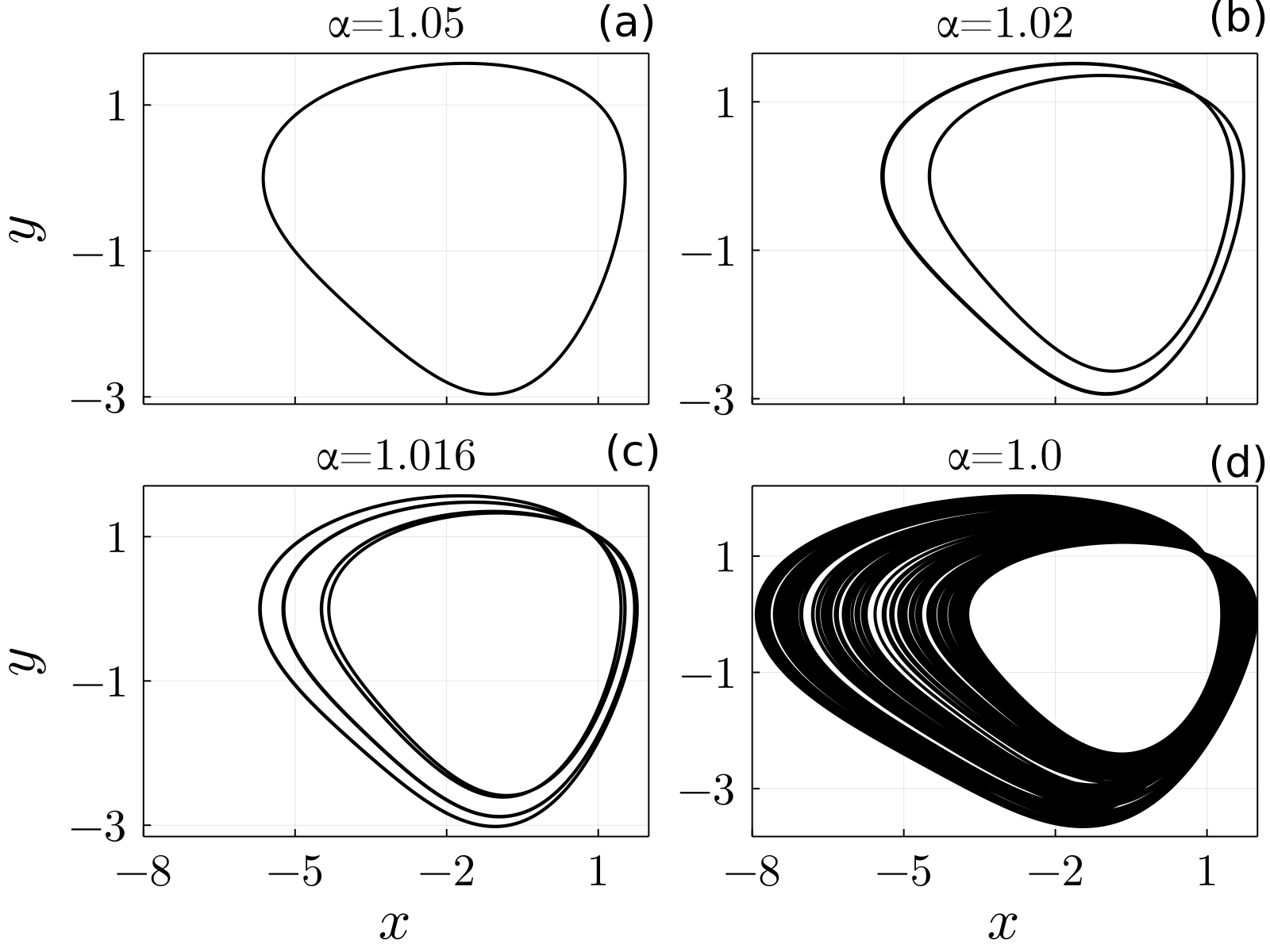}}} 
	\caption{ (a) Period-1 ($\alpha=1.05$) (b) period-2 ($\alpha=1.02$), (c) period-4 ($\alpha=1.016$) and (d) chaotic ($\alpha=1$)  attractors for system, Eq. (\ref{wang}).}
 \label{fig:wang_phase}
 \end{figure}

{\it{Hidden motion without any fixed point:}} Consider another dynamical system with hidden attractors \cite{Jafarinoeq}, which has no equilibrium point:
\begin{equation}
\dot{x}=y,~~~ \dot{y}=z, ~~~\dot{z}=-y+\gamma x^2+\beta x z +\alpha,
\label{wang}
\end{equation}
\noindent where $\alpha$, $\beta$, and $\gamma$ are parameters. In general, the system has two fixed points $(\pm\sqrt{-\alpha/\gamma},0,0)$ for any one of the parameters $\gamma<0$ or $\alpha<0$. The system has no equilibria if both of these parameters are positive. We fix the parameters $\beta=1.1$ and $\gamma=0.1$ and vary $\alpha$ from 1 to 1.06, i.e., there is no fixed point for this range. Figure \ref{fig:wang} shows the orbit diagram overlapped with the continuation diagram of the system in Eq. (\ref{wang}) as a function of parameter $\alpha$. It shows that a periodic orbit of large amplitude is created near $\alpha\sim 1.058$. A period-doubling bifurcation leads to chaotic motion as the parameter $\alpha$ is decreased.
Fig. \ref{fig:wang_phase} shows the trajectory in phase space of the system Eq. (\ref{wang}) for various values of the parameter $\alpha$. Fig. \ref{fig:wang_phase}(a) shows the period-1 orbit for $\alpha=1.05$, \ref{fig:wang_phase}(b) is the periodic-2 limit cycle for $\alpha=1.02$, \ref{fig:wang_phase}(c) shows the period-4 limit cycle for $\alpha=1.016$ and \ref{fig:wang_phase}(d) is the chaotic attractor for the value of $\alpha=1.0$. Note that all these attractors are hidden attractors. 
Unlike the previous case, Eq. (\ref{wang}) has no fixed point and hence the system has no global attractors \cite{nkv}. Euclidean phase space cannot have a global attractor without any equilibrium states because a global attractor must attract all orbits, which is not possible without any fixed points or equilibrium states. This is due to the fact that a fixed point is necessary for convergence of nearby orbits, which is required for the existence of a global attractor \cite{Arnold1}. 
For $\alpha>1.058$, the system does not have basin boundaries and attractors as there are no local bifurcations. However, when the parameter is decreased to $\alpha<1.058$, the system exhibits a saddle-node bifurcation on orbit (SNO) which leads to the birth of hidden attractors. The stable and unstable periodic orbits are represented by SPO and UPO, respectively, in Fig. \ref{fig:wang}. This global bifurcations (SNO) create basin boundaries that have attractors in the system. In the interval of $\alpha \in(1,1.058)$, the system has a basin of attraction for the hidden attractor created at the saddle-node bifurcation. As $\alpha$ decreases further, the hidden attractor disappears as it collides with the unstable periodic attractor. Note that the hidden attractor exists only within a small basin created by the saddle-node bifurcation, and initial conditions outside the basin diverge to infinity.
\par {\it{Experimental evidence:}} 
 The existence of hidden attractors in nonlinear electronic circuits is demonstrated in this section. The design of such circuits can be realized through the utilization of coupled first-order differential equations. However, identifying hidden attractors within these circuits posses a challenging task that requires a careful experimental approach. To detect these attractors, the appropriate initial conditions must be selected from the appropriate basin of the attraction.
\begin{figure}
  \centering
 {\scalebox{0.45}{\includegraphics[angle=0]{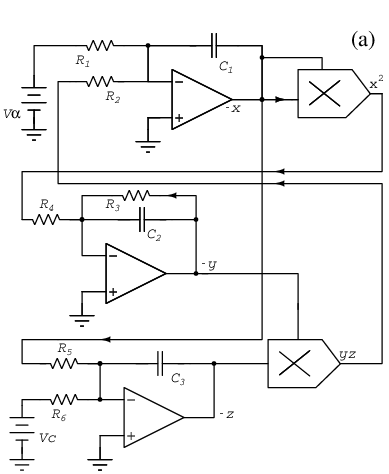}}} 

 {\scalebox{0.3}{\includegraphics{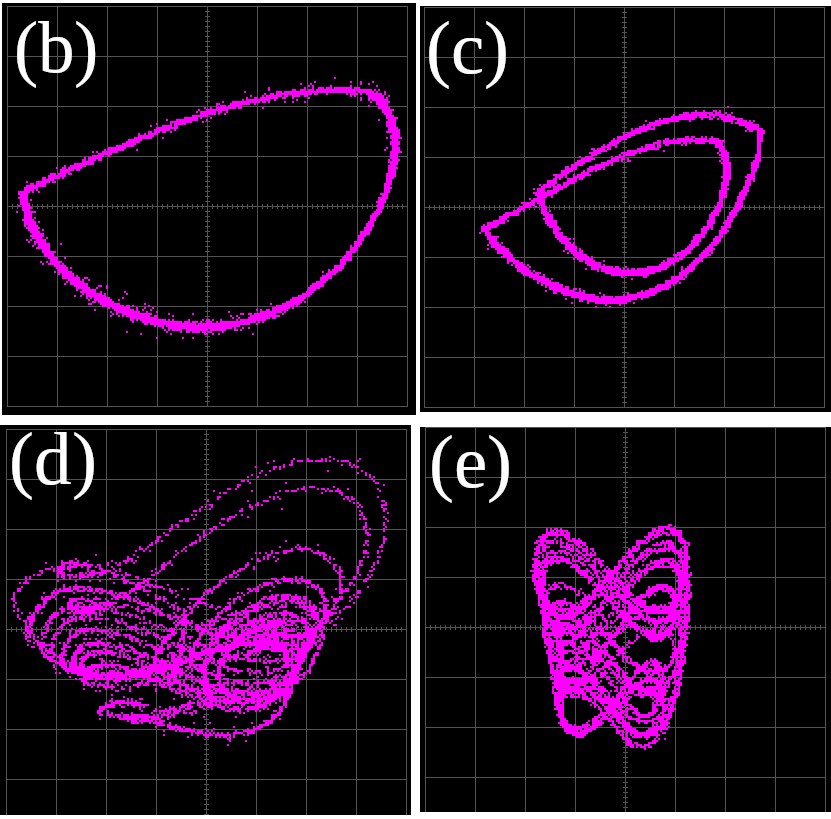}}}
 \caption{Experimental result: (a) circuit diagram for the Eq. \ref{wang1}.  (b) Period-one limit cycle for $v_\alpha=0.075$, (c) Period-two for the $v_\alpha=0.07$, (c) and (d) are the chaotic attractor for $v_\alpha=0.068$ and $v_\alpha=0.002$ respectively.}
 \label{fig:wang_circuit}
 \end{figure}

The present study demonstrates the detection of hidden attractors in nonlinear electronic circuits using $\mu$A741 ICs for the integrator part of the circuit and AD633AN ICs for the multiplier function. The resistors used in the circuit were selected as $R_1=R_3=R_6=100k\Omega$, $R_2, R_4=10k\Omega$, and $R_5=25k\Omega$. The capacitors in the integrator were valued at $C_1=C_2=C_3=10nF$.

The results of the circuit design are presented in Fig. \ref{fig:wang_circuit}(a). As the parameter corresponding to $\alpha$ ($v_\alpha$), is decreased from 0.1V, the circuit shows a periodic doubling route to chaos. The phase space corresponding to the period doubling route chaos is presented in Figs. \ref{fig:wang_circuit}. Figs \ref{fig:wang_circuit}a \& \ref{fig:wang_circuit}b show the period-one and period-two oscillations for the parameter $v_\alpha=0.075$ and $v_\alpha=0.07$ respectively. Figs.  \ref{fig:wang_circuit}c \& \ref{fig:wang_circuit}d are the chaotic attractors for two different values of $v_\alpha=0.068$ and $v_\alpha=0.002$ respectively.

In summary, in this study, we demonstrate the existence of hidden attractors in nonlinear dynamical systems.  We demonstrate that hidden attractors are a consequence of a saddle-node bifurcation of orbits, which is the result of a collision between stable and unstable periodic orbits. We explore two different scenarios: (i) systems with one stable equilibrium point that contain hidden attractors, and (ii) systems with no equilibrium points. In the first scenario, hidden attractors arise due to the loss of global stability and the occurrence of global bifurcations, specifically saddle-node bifurcations of periodic orbits, leading to the creation of basins with no associated equilibrium points. In the second scenario, the absence of equilibrium points results in a lack of global stability or global attractors, leading to unstable solutions. By selecting suitable parameters, we observe the emergence of hidden attractors as a result of the collision between stable periodic and unstable periodic orbits, which is a saddle-node bifurcation of orbits. We have constructed real-time electronic circuits to demonstrate the system with one equilibrium point and the experimental results support our findings and verify the presence of hidden attractors in these systems. We found similar results in various systems with hidden attractors, regardless of whether they contain one stable equilibrium point or no equilibrium points. Although we have not presented results for general dynamical systems with hidden attractors, the results we have presented answer one of the intriguing questions related to the emergence of hidden attractors in nonlinear dynamical systems.

SK acknowledge the Center for Nonlinear Systems, Chennai Institute of Technology (CIT), India, vide funding number CIT/CNS/2023/RP-016. MDS and NVK would like to acknowledge the funding support by DST-RSF, Indo-Russia project vide grant no. INT/RUS/RSF/P-18/2019. AP
thanks IoE, the University of Delhi for financial assistance.

\end{document}